\documentclass{article}
\usepackage[utf8]{inputenc}
\usepackage{amsmath}
\usepackage{amssymb}
\usepackage{parskip}
\usepackage{graphicx}
\graphicspath{{./pic/}}
\usepackage[a4paper, total={6in, 10in}]{geometry}
\usepackage[font=small,labelfont=bf]{caption}
\usepackage{tabularx}
\newcommand{\RNum}[1]{\uppercase\expandafter{\romannumeral #1\relax}}
\title{An integrated vertex model of the mesoderm invagination during the embryonic development of Drosophila}
\author{Jianfei Jiang and Christof M. Aegerter}

\begin{document}

\maketitle

\section{Abstract}
The mesoderm invagination of the Drosophila embryo is known as an archetypal morphogenic process. To explore the roles of the active cellular forces and the regulation of the forces, we proposed an integrated vertex model that combines the regulation with the cell movements. Our results suggest that a successful furrow formation requires an apical tension gradient, decreased basal tension, and increased lateral tension, which corresponds to apical constriction, basal expansion, and apicobasal shortening respectively. Our model also considers the mechanical feedback which leads to an ectopic twist expression with external compression as observed in experiments. Our model predicts that ectopic invagination could happen if the external compressive gradient is applied.

\section{Introduction}
Plenty of research has been conducted to investigate the mechanisms of the mesoderm invagination of the Drosophila embryo. The biological basis of this morphogenic process is well studied including the detailed genetic control network, generation of cellular forces, and tissue-wide force propagation, which is reviewed systematically \cite{martin2020physical}.The gist is that the morphogen gradient sets up a series of gene expressions that pattern a graded distribution of the apical myosin \RNum{2} intensity \cite{heer2017actomyosin}.The cellular force generated by myosin motors on the apical surface is also graded, which is measured through video force microscopy (VFM) \cite{Brodland22111}.The apical force gradient promotes the contraction of the apical surface and from an inward curvature. This tissue shape change is so-called apical constriction which predominantly happens in the mesoderm. With the help of the radial shortening, a closed furrow forms with a certain depth. There are a lot of mathematical models describing the possible mechanisms of the mesoderm invagination. For example, the inverse algorithm of VFM provides a mathematical model that calculates the cell movements with a given set of force distribution. The model suggests that the apical constriction resulting from the apical force gradient promotes the formation of inward curvature and has little effect on the furrow depth. In addition, apicobasal shortening is thought to be the decisive mechanism of the invagination depth \cite{conte2012biomechanical}.This is supported by the experiment \cite{gracia2019mechanical}, in which disruption of the apical-basal forces stops the mesoderm invagination. In contrast, the passive mechanical model \cite{polyakov2014passive} suggests that the apical constriction and basal expansion are sufficient to replicate the mesoderm invagination. The model regards the apical constriction as the only active mechanism. The basal expansion results from the decrease of the basal stiffness which might be attributed to the depletion of basal myosin \RNum{2} intensity observed in vivo \cite{polyakov2014passive}.Another vertex model proposed by Hočevar Brezavšček et al \cite{brezavvsvcek2012model} aims to explore the minimal requirements of the mesoderm invagination and concludes that the invagination could be a result of collective instability although the assumption has not been verified experimentally.
\\
\\
We would like to study the roles of the active forces during the mesoderm invagination, as well as the regulatory network controlling the dynamics of the forces. To this end, we propose a modified vertex model concerning both active tensions and passive mechanical responses. We examine our model with a given set of static tensions and tensions with prescribed dynamics. We demonstrate that the mesoderm invagination requires the radial shortening force, namely lateral tension, which affects the furrow depth. And high basal tension resists internalization. We also show that with dynamic tensions, apical constriction cannot promote the formation of a closed ventral furrow alone, but together with apicobasal shortening and basal expansion, the mesoderm invagination proceeds and eventually forms a tubular furrow. In addition, we come up with a concise regulation network by setting up reaction-diffusion equations to describe possible tension dynamics with the consideration of mechanical feedback. The integration of the regulatory network and vertex model also yields a successful invagination. And we test the model with external constraints, i.e. global compression and local compressive gradient. We show that global compression induces the ectopic expression of twist which flattens the apical force gradient and fails to invaginate normally. The local compressive gradient sets up an ectopic twist gradient, thus, force gradient, and leads to the ectopic invagination.

\section{Methods}
\begin{figure}[h]
    \centering
    \includegraphics[width=0.5\textwidth]{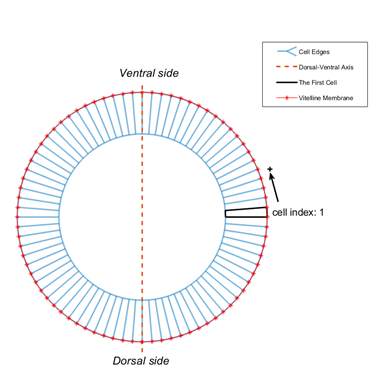}
    \caption{Demonstration of initial setup of the vertex model. Cell index increases counterclockwise as shown by the black arrow. The number of cells in the tissue is 80.}
    \label{fig:1}
\end{figure}

In our mechanical model, the tissue consists of individual cells described by quadrilaterals as previous models \cite{brezavvsvcek2012model,polyakov2014passive} (Figure \ref{fig:1}). We consider both active forces and passive mechanical responses in our model by assigning active forces to each side of the cell and having an area elasticity term. In contrast to the previous model, we allow cells and the yolk to have small elastic deformation, which is similar to Farhadifar’s work \cite{farhadifar2007influence}.Then potential energy of the entire tissue is given by Equation \ref{eq:1}.
\begin{equation} \label{eq:1}
    E(\boldsymbol{r})=\sum_{i=1}^N[\alpha L_a^i(\boldsymbol{r})+\beta L_b^i(\boldsymbol{r})+\frac{1}{2}\gamma L_l^i(\boldsymbol{r})+\frac{1}{2}K_a(A_i \boldsymbol{r}-A_0 )^2]+\frac{1}{2}K_y(A_y (r)-A_y0 )^2
\end{equation}

\(\alpha\),\(\beta\) and \(\gamma\) are the line tension coefficients of the apical, basal and lateral sides; \(L_a^i\) and \(L_b^i\) are lengths of apical and basal sides, \(L_l^i\) is the sum of two lateral lengths of \(i\)-th cell. \(A\)  is the area, subscript \(y\) denotes the yolk, subscript 0 denotes the preferred area; \(K_a\) and \(K_y\) are the elasticity constants of the cell and the yolk respectively. The vitelline membrane applies an extra energy term  \ref{eq:2} to the vertices outside the membrane depending on the distance from the membrane.
\begin{equation} \label{eq:2}
    E_{ext}(\boldsymbol{r}) = E_{press}(\exp(\frac{r-r_0}{r}) - 1)    \;\; for \;\; r > r_0
\end{equation}

\(E_{press}\) is the amplitude of the extra energy provided by the membrane. Equation \ref{eq:2} is set to zero for vertices inside the membrane.\(r_0\) is the radius of the membrane. Another additional energy term is considered for invaginated cells, which is the adhesion term. This term minimizes the distance between the vertices that are symmetric about the ventral midline to help form a closed furrow.

\begin{equation} \label{eq:3}
    E_{ad} = 
\begin{cases}
    Ad & \text{if $d<d_0$}
   \\ 0 & \text{else}
\end{cases}
\end{equation}
\(d\) is the distance between the symmetrically distributed vertices, \(d_0\) is the distance that adhesion occurs. \(Ad\) is constant adhesion energy. The exact expression of adhesion is not important, the point is that the adhesion term makes invaginated cells stick to each other rather than overlap. Based on experiments \cite{Brodland22111,heer2017actomyosin}, the apical constriction is driven by apical tension gradient. We use a spatially dependent apical line tension coefficient, i.e. a Gaussian distribution.
\begin{equation} \label{eq:4}
    \alpha(n) = \alpha_0\exp(-\mu (n - n_0)^2) + \alpha_1
\end{equation}
where \(\mu\) defines the mesoderm size, \(n_0\) is the cell index of the ventral midline, \(\alpha_1\) is the overall level of the apical tension, \(\alpha_0\) is the amplitude. The net force of each vertex is computed through the energy gradient and the motion is thought to be overdamped \cite{drasdo2000buckling,fletcher2013implementing}. Due to the small spatial scale, the acceleration term is much smaller than the dissipative term and could be neglected, leading to the equation of motion of vertices:
\begin{equation} \label{eq:5}
    \eta \frac{d\boldsymbol{r}_i}{dt} = \boldsymbol{F}_i
\end{equation}
where \(\eta\) is the drag factor. Note that the parameters used for the simulations are pure numbers, the actual values of the quantities should scale with the parameters. The units do not matter the results, but the relations between parameters do.

\section{Results}
\subsection{Static tensions}
\begin{figure}[h]
    \centering
    \includegraphics[width=\textwidth]{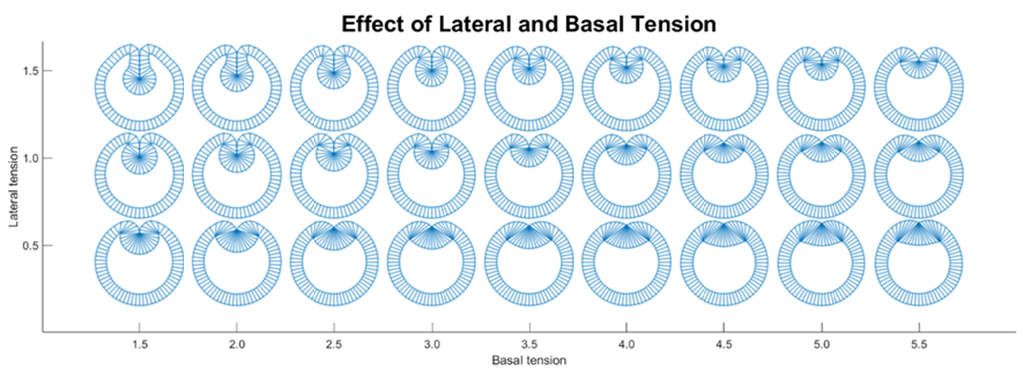}
    \caption{Phase diagram of fixed apical tension}
    \label{fig:2}
\end{figure}

First, we fixed the apical tension and vary basal and lateral tension within a proper domain. For simplicity, we assumed that no spatial dependence in basal and lateral tensions in this part. We obtained a phase diagram (Figure \ref{fig:2}) showing the effects of basal and lateral tensions. Parameters used in the simulations are shown in Table 1. We choose lateral tension to vary around a small level so that our model works properly. This management is reasonable according to the VFM measurements, where the measured lateral forces seem to be much smaller than basal and apical forces \cite{Brodland22111}. We can see that along the horizontal axis, raising basal tension flattens the internalized curvature, in contrast, along the vertical axis, the furrow gets deeper if the lateral tension is increased. Our simulations exhibit similar behavior as Conte’s model \cite{conte2012biomechanical} where apical constriction hardly matters the furrow depth, and the critical factor governing the internalization is the radial shortening force. Although we did not change the apical gradient, our model shows that the increase of the lateral tension amplifies the radial shortening, thus increasing the furrow depth. A closed furrow also requires small basal tensions. From the bottom right to the top left of Figure \ref{fig:2}, the tissue undergoes apical constriction due to the force gradient, apicobasal elongation happens associated with the contraction due to a passive cellular response; then, since the basal tension drops and lateral tension grows, the basal expansion, as well as apicobasal shortening, happens cooperatively leading to the formation of a closed furrow. This may explain the dynamic process of mesoderm invagination. In comparison to Polyakov’s model \cite{polyakov2014passive}, in which the basal expansion is achieved by the reduction of the basal stiffness, our model suggests that the depletion of active basal force without affecting the tissue mechanical properties could lead to a similar effect. The observed decrease of the basal myosin intensity \cite{polyakov2014passive} Polyakov’s model shows that apical constriction and passive mechanical responses are sufficient to drive mesoderm invagination, nevertheless, our model emphasizes the roles of active basal and lateral forces.
\begin{table}[]
    \centering
    \begin{tabularx}{\textwidth}{ 
    | >{\centering\arraybackslash}X 
    | >{\centering\arraybackslash}X 
    | >{\centering\arraybackslash}X
    | >{\centering\arraybackslash}X
    | >{\centering\arraybackslash}X
     | >{\centering\arraybackslash}X| }
     \hline
     $\alpha_0$ & 9 & $A_0$ & 0.0314 & $Ad$ & 10\\
     \hline
     $\alpha_1$ & 1 & $K_y$ & $10^2$ & $d_0$ & 0.0471\\
     \hline
     $\mu$ & 0.02 & $A_{y0}$ & 2.0086 & $\eta$ & 1 \\
    \hline
     $n_0$ & 20.5 & $E_{press}$ & 10 & dt & $1.5\times 10^{-4}$\\
     \hline
     $K_a$ & $10^4$ & $r_0$ & 1.2 & N & $10^5$\\
     \hline
    \end{tabularx}
    \caption{cell-specific parameters and iteration variables. N is the total number of iterations.}
    \label{tab:1}
\end{table}

\subsection{Dynamic tensions}
\begin{figure}[h]
    \centering
    \includegraphics[width=0.5\textwidth]{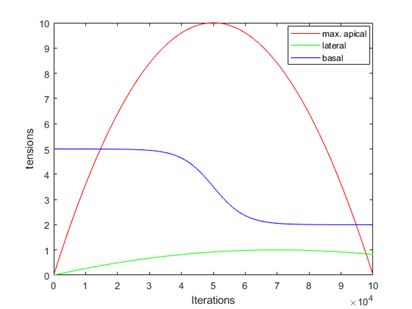}
    \caption{prescribed time-dependence of tensions. It mimics the establishment and relaxation of apical tension gradients, the increase of lateral tension and the basal force depletion}
    \label{fig:3}
\end{figure}

\begin{figure}[h]
    \centering
    \includegraphics[width=\textwidth]{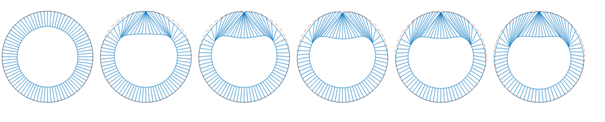}
    \caption{Shapes of different stages during simulations and are in time-order. The time interval between successive frames is 20\% of the total simulation time.}
    \label{fig:4}
\end{figure}

\begin{figure}[h]
    \centering
    \includegraphics[width=\textwidth]{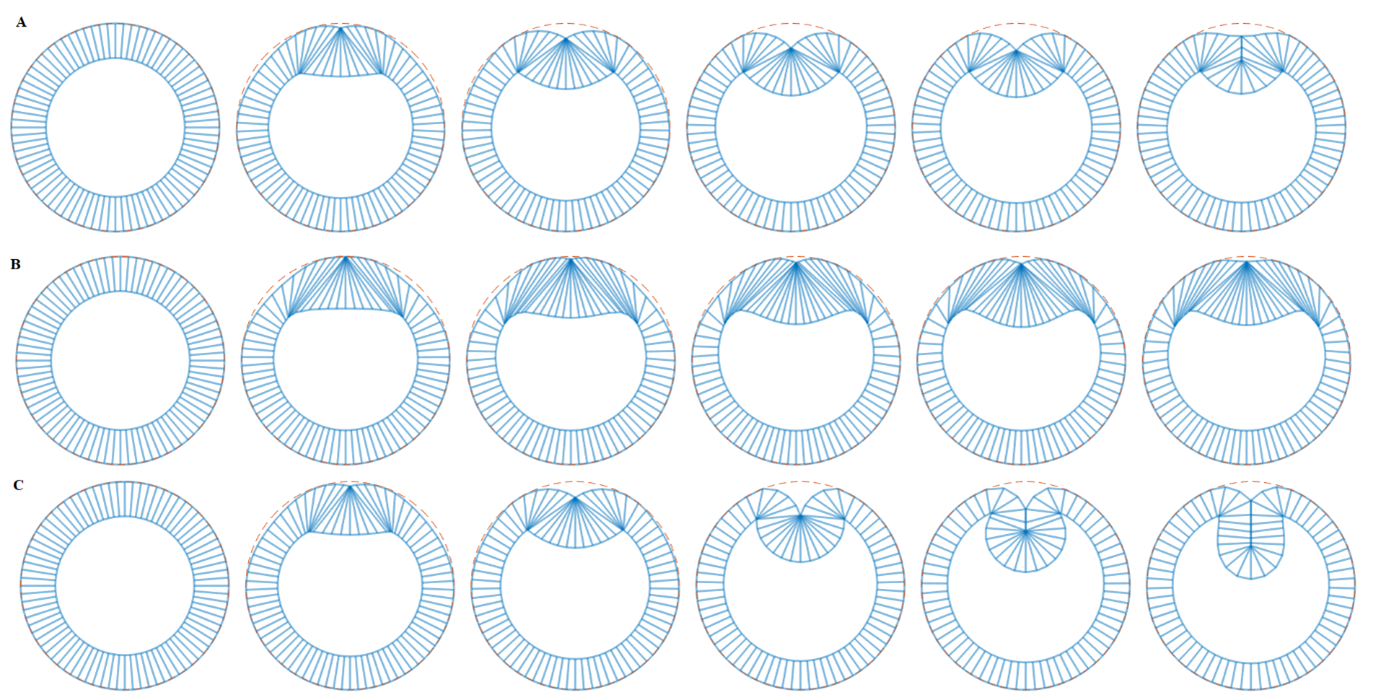}
    \caption{(A) apical constriction \& apicobasal shortening. Without basal depletion, insufficient internalization happens. (B) apical constriction \& basal expansion. Lack of apicobasal shortening fails to activate further internalization. (C) apical constriction, basal expansion \& apicobasal shortening. A tubular furrow forms in presence of the three mechanisms. The time interval between successive shapes is 20\% of the total simulation time.}
    \label{fig:5}
\end{figure}

Active forces are probably time-dependent according to VFM measurements \cite{Brodland22111} No matter what the mechanisms regulating the dynamics are, we applied the dynamic tension whose time-dependence is prescribed by Equation \ref{eq:6} and \ref{eq:7} to our model based on the VFM results. 
\begin{equation} \label{eq:6}
    f(t) = -f_{max}/t_m^2 (t - t_m)^2  + f_{max}
\end{equation}
\begin{equation} \label{eq:7}
    f(t) = \frac{A}{\exp(\kappa(t-t_0))+1} + B
\end{equation}
Equation \ref{eq:6} is a parabolic function to describe the maximum of apical tension and overall level of lateral tension. The depletion of basal tension is described by a sigmoid function Equation \ref{eq:7}. In the above equations, \(f_max\) is the maximum of the parabolic function, \(t_m\) is the time point of the maximum, \(A\) is the amplitude of the sigmoid function, \(B\) is the baseline, \(\kappa\) describes the decreasing rate and \(t_0\) is the time point of half-amplitude reduction. We still assumed that there is no spatial dependence on the basal and lateral tension. Coefficients used for simulation are given in Table 2. The resulting time dependences of the tensions are shown in Figure \ref{fig:3}.
\begin{table}[h]
    \centering
    \begin{tabular}{|c|c|c|c|c|c|c|c|c|} 
    \hline
        $f_{max_{apical}}$ & $t_{m_{apical}}$ & $f_{lateral}$ & $t_{m_{lateral}}$ & $A$ & $B$ & $\kappa$ & $t_0$  \\
        \hline
        10 & 0.5N & 1 & 0.7N & 3 & 2 & $\frac{20}{N}$ & 0.5N\\
        \hline
    \end{tabular}
    \caption{parameters for tension dynamics. N is the total number of iterations and is the same as in Table \ref{tab:1}}
    \label{tab:2}
\end{table}
With prescribed tension dynamics, we now have three mechanisms governing the mesoderm invagination: apical constriction, apicobasal shortening, and basal expansion. To study their contributions, we examined different combinations of those mechanisms. First, only apical constriction is considered. The resulting shapes in the time series are shown in Figure \ref{fig:4} . The results suggest that apical constriction cannot promote the formation of a closed furrow alone, which is consistent with the conclusion of Conte’s model \cite{conte2012biomechanical}.
Then we added basal force depletion and lateral force increase separately to apical constriction. The results demonstrate that neither of them can lead to a successful invagination unless all three mechanisms are involved. (Figure \ref{fig:5})
Our model based on apical force gradient shows the necessity of basal expansion and apicobasal shortening. A recent study confirms that apical-basal force is required for the mesoderm invagination, the disruption of that force stops the invagination \cite{gracia2019mechanical}. Basal myosin depletion is also observed experimentally \cite{polyakov2014passive}.
\subsection{Integrated with regulatory network}
\begin{figure}[h]
    \centering
    \includegraphics[width = 0.75\textwidth]{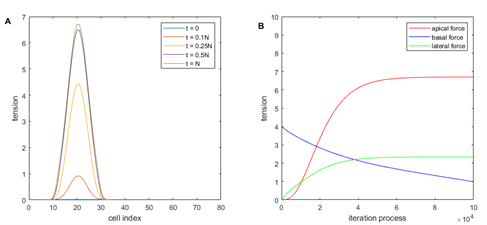}
    \caption{the resulting tension dynamics. (A) spatial-resolved apical tension. (B) time dependencies of forces at the ventral midline.}
    \label{fig:6}
\end{figure}

We proposed a regulatory network to account for the tension dynamics. Since the entire genetic pathway is complicated, a feasible way is to neglect intermediate factors. The network only contains one transcription factor \emph{twist} which is thought to have a graded distribution and affect the myosin \RNum{2} translocation in the cell. Assume that \emph{twist} has a positive effect on apical and lateral myosin \RNum{2} concentration and a negative effect on basal myosin \RNum{2} concentration. We started with Gierer-Meinhardt equation \cite{gierer1972theory} used for \emph{twist} as an activator:
\begin{equation} \label{eq:8}
    \frac{dC_{\emph{twi}}}{dt} = \frac{p_{twi}C_{twi}^2}{1+K_{twi}C_{twi}^2} - \mu_{twi}C_{twi} + p_0\exp(-\mu(n-n_0)^2)
\end{equation}
And myosin \RNum{2} concentrations are governed by:
\begin{equation}\label{eq:9}
    \frac{dC_{Amyo}}{dt} = p_AC_{twi}^2 - \mu_{myo}C_{Amyo}
\end{equation}
\begin{equation}\label{eq:10}
    \frac{dC_{Bmyo}}{dt} = -lC_{twi}^2 +p_BC_{Bmyo}- \mu_{myo}C_{Bmyo}^2
\end{equation}
\begin{equation}\label{eq:11}
    \frac{dC_{Lmyo}}{dt} = p_LC_{twi}- \mu_{myo}C_{Lmyo} + p_{L0}
\end{equation}
In above Equations, \(p\) denotes the production rate of the substance specified by the subscript, \(C_i\) is the concentration of substance \(i\), \(\mu_{myo}\) is the myosin breakdown rate. To simulate the myosin depletion, we use a logistic growth in Equation \ref{eq:10} for the regulation of the basal myosin, thus the breakdown rate \(\mu_B\) differs from \(\mu_{myo}\). Equation \ref{eq:9} and \ref{eq:11} are sufficient to mimic the increase of apical and basal myosin intensity. The constant production term \(p_{L0}\) of Equation \ref{eq:11} is expected to account for the increase of lateral myosin intensity. Translating concentration to force is achieved by multiplying a factor \(f\) which could be tension-specific, but in our parameter set, \(f\) is set to 1 for all tensions. Other parameters are given in Table \ref{tab:3}, and we obtained the time dependencies of the tensions shown in Figure \ref{fig:6}.
\begin{table}[h]
    \centering
    \begin{tabular}{|c|c|c|c|c|c|c|c|c|c|c|c|c|c|} 
    \hline
        $p_{twi}$ & $\mu_{twi}$ & $p_0$ & $K_{twi}$ & $\mu$ & $p_A$ & $\mu_{myo}$ & $l$ & $p_B$ & $\mu_B$ & $p_L$ & $p_{L0}$ & $f$  \\
        \hline
        1 & 0.8 & 1.5 & 1 & 0.02 & 0.75 & 1 & 0.03 & 0.25 & 0.1 & 0.15 & 1 & 1\\
        \hline
    \end{tabular}
    \caption{parameters for the regulatory equations}
    \label{tab:3}
\end{table}

So far, the regulatory equations manage to replicate the apical tension gradient, basal tension decrease, and lateral tension increase. In addition, the mechanical information could affect the twist production as shown by Farge’s experiment \cite{farge2003mechanical}. To take such mechanical effects into account, we modify the last term of Equation \ref{eq:8} as follows.
\begin{equation} \label{eq:12}
    \frac{dC_{\emph{twi}}}{dt} = \frac{p_{twi}C_{twi}^2}{1+K_{twi}C_{twi}^2} - \mu_{twi}C_{twi} + p_0\frac{P^q}{P^q + H^q}
\end{equation}
\begin{equation} \label{eq:13}
    P = a\exp(-\mu (n-n_0)^2) + b\frac{\Delta L^l}{\Delta L^l + H_m^l}
\end{equation}
The production term in Equation \ref{eq:12} is given by Equation \ref{eq:13} in which a Gaussian distribution governs the intrinsic spatial dependence, and a Hill function governs the mechanical response on \emph{twist} production. \(\Delta L\) is the deformation of the apical length relative to its rest value \(L_0\). \(a\) and \(b\) are the contributing factors. \(H\) and \(H_m\) are the values when the corresponding Hill function reaches its half-saturation. Both genetic and mechanical terms share the same saturation of the Hill function in Equation \ref{eq:12} to avoid an extremely high production rate at the ventral side where both genetic and mechanical terms are very high. The genetic and mechanical terms are treated in the same manner, and the bias is only given by the contribution factor. Parameters for the modified equations are given in Table \ref{tab:4}, unmentioned parameters are the same as Table \ref{tab:3}.
\begin{table}[h]
    \centering
    \begin{tabular}{|c|c|c|c|c|c|c|}
    \hline
        $q$ & $H$ & $l$ & $L_0$ & $H_m$ & $a$ & $b$  \\
        \hline
        3 & 0.3 & 8 & 0.0942 & 0.1 & 0.5 & 0.5\\ 
        \hline
    \end{tabular}
    \caption{parameters of modified regulatory equations.}
    \label{tab:4}
\end{table}
Note that high Hill coefficient \(l\) in Equation \ref{eq:13} enables a rapid switching behavior for the mechanical response so that the mechanical information triggers a part of the \emph{twist} production. In Equation \ref{eq:12}, the Hill coefficient q is much smaller, which allows a smoother transition from low production to saturation. 
\\
\\
\subsubsection{Normal invagination}
\begin{figure}[h]
    \centering
    \includegraphics[width=\textwidth]{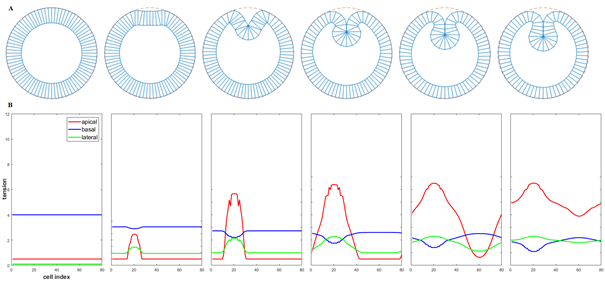}
    \caption{(A) normal invagination process in time order. The time interval between successive frames is 20\% of the total simulation time. (B) the corresponding tension profiles of shapes in (A). The red line is apical tension, the blue line is basal tension, and the green line is lateral tension.}
    \label{fig:7}
\end{figure}

Our model can reproduce the normal invagination process, which ends up with a closed ventral furrow. (Figure \ref{fig:7}A) However, the intermediate shapes do not strictly coincide with the experimental observation. The difference may result from the early activation of the apicobasal shortening due to the increase of the lateral tension (Figure \ref{fig:7}B). The mesoderm bends inward rather than undergoing apical constriction and apicobasal shortening in sequence. The abnormal apical constriction might be attributed to the mechanical feedback term in Equation \ref{eq:12}.That term acts as positive feedback in contracted cells and negative feedback in stretched cells. Therefore, at later stages in our model, the apical tension in the ectoderm is raised to maintain the apical length round the rest length. The apical tension gradient is flattened due to the mechanical feedback. However, the flattened gradient does not stop the mesoderm invagination from proceeding, suggesting that once the internalized curvature has formed, the apicobasal shortening force plays a more important role for the furrow formation, which is, again, consistent with the previous model \cite{conte2012biomechanical}. Despite the deviation from the in-vivo observation, our integrated model can successfully invaginate in presence of mechanical feedback.
\subsubsection{Elliptic compression}
\begin{figure}[h]
    \centering
    \includegraphics[width = \textwidth]{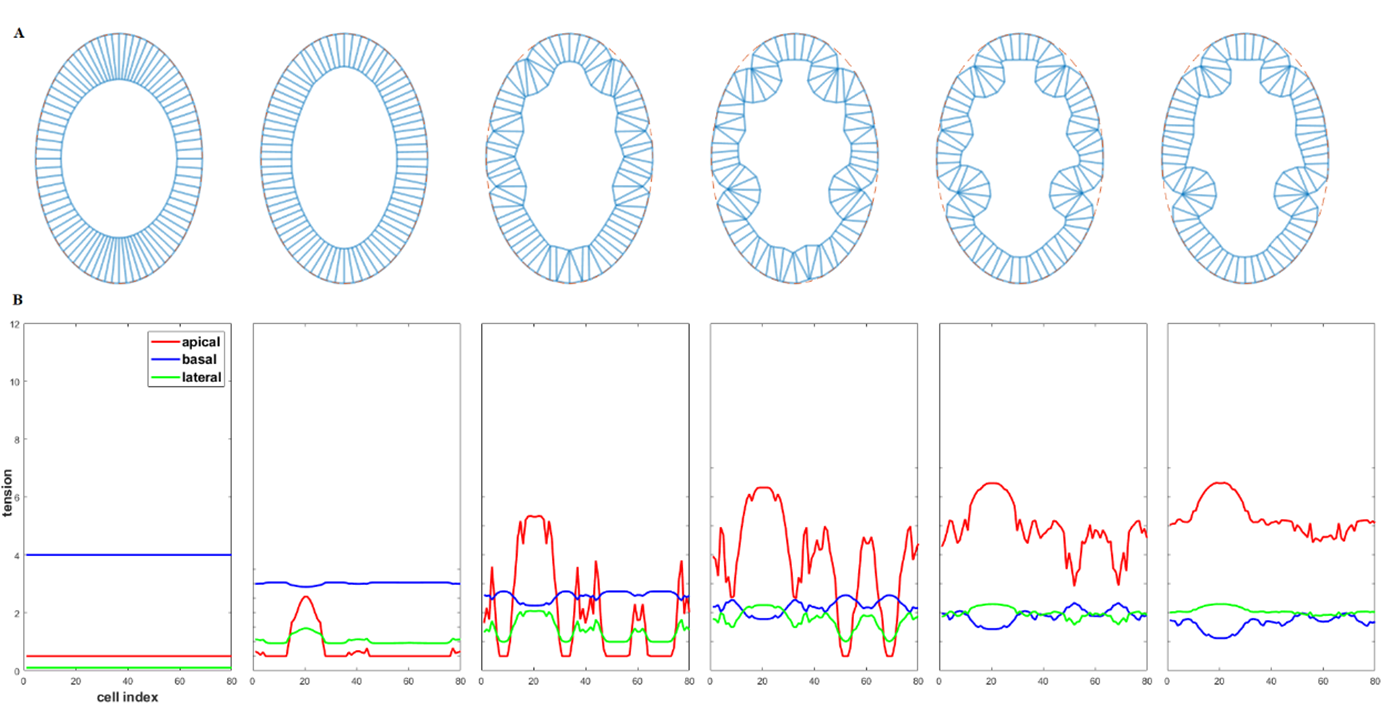}
    \caption{elliptic compression. The time interval between successive frames is the same as the previous simulation. (A) Tissue shape change (B) Tension profiles of corresponding shapes.}
    \label{fig:8}
\end{figure}

To study the influences of the mechanical feedback term in Equation \ref{eq:13} , we examined our model under different mechanical constraints. For instance, according to Farge’s experiment \cite{farge2003mechanical} expression of the \emph{twist} gene could be mechanically induced by compressing the embryo. In the experiment, the global compression is achieved by pressing the coverslip above the embryo, leading to an elliptic-shaped outline of the embryo. To simulate this compression, we changed our contour of the vitelline membrane to be an ellipse, with 20\% extension along the dorsal-ventral axis and 20\% compression in the transversal direction to the dorsal-ventral axis. We perform the energy minimization to find the initial equilibrium for the new boundary. Then the simulation started from that elliptic initial shape.
\\
\\
It turns out that the compressed tissue does not proceed with mesoderm invagination anymore, although there are anomalous apical constrictions along with the tissue. (Figure \ref{fig:8}) The anomalies are located symmetrically along the circumference of the embryo. This is because cells are contracted apically at those positions due to the elliptic compression, the contractions exceed the threshold and switch on the positive feedback of twist production, leading to apical constrictions. Around the dorsal-ventral axis, the apical sides of the cells are stretched, the mechanical feedback is negative such that the lengths remain within a certain tolerance. The \emph{twist} gradient is flattened by the mechanical feedback term leading to a smoother apical force distribution, which is unable to create an inward curvature. Therefore, even having increased lateral force, the invagination cannot happen. 
In addition, the tension profiles are oscillatory, which may result from the even number of the Hill coefficient in Equation \ref{eq:13}. The mechanical term is sensitive to the absolute value of deformation, either compression or extension can induce \emph{twist} production, which is also the reason why the feedback term behaves differently in stretched and compressed cells. Despite the abnormal apical constrictions, our model reproduces the phenomenon that ectopic \emph{twist} production induced mechanically fails the mesoderm invagination. The simplicity of our regulatory network may account for the anomalies. There could be other mechanisms and regulatory rules to avoid those unexpected deformations.
\subsubsection{Local ectopic compression}
\begin{figure}[h]
    \centering
    \includegraphics[width = 0.5\textwidth]{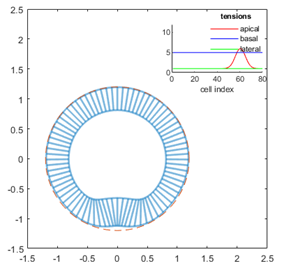}
    \caption{Initial local compression on the ectopic side with external tensions shown in the inset. 5000 iterations are used to obtain this shape.}
    \label{fig:9}
\end{figure}

\begin{figure}[h]
    \centering
    \includegraphics[width = \textwidth]{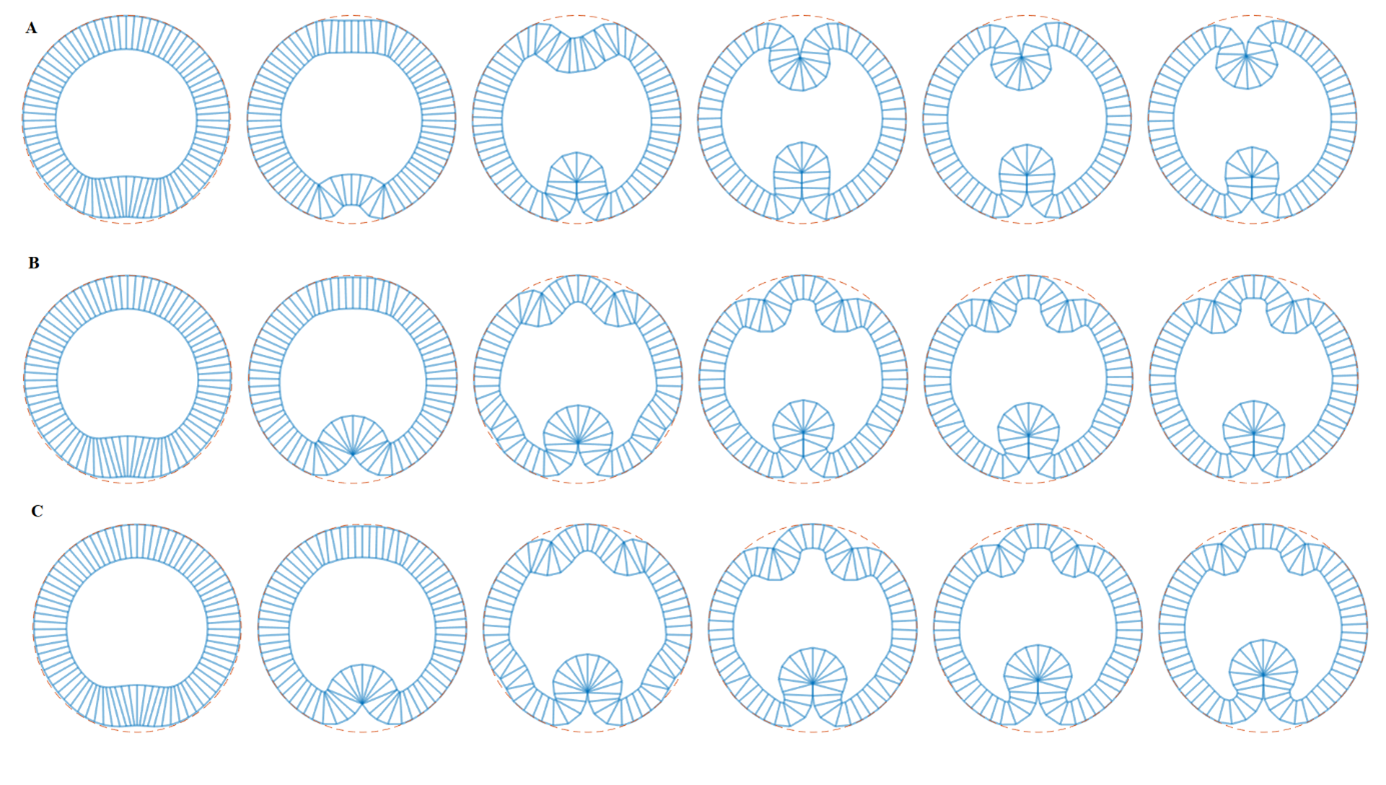}
    \caption{(A) Initial ectopic compression without holding the external force. (B) hold the external force for half of the total simulation time. (C) hold external force for the entire simulation time.}
    \label{fig:10}
\end{figure}

Furthermore, we investigated the behaviors of our model under external force gradient. We applied the external force gradient, which peaks at the dorsal midline, to the ectoderm in order to get a deformation gradient. We held the external force for a certain amount of time while executing the energy minimization algorithm to get the initial shape (Figure \ref{fig:9}) and started the simulation from that shape. We retreated the external force at different time points to see whether ectopic invagination could happen with such a mechanical constraint. It turns out that our model predicts the occurrence of ectopic invagination no matter when the external force is switched off. Holding the external force for a long time tends to stop mesoderm invagination (Figure \ref{fig:10}).

Our model shows that the initial deformation gradient on the ectoderm could mechanically induce the twist gradient which sets up an ectopic force gradient. This force gradient amplifies the internalization of the ectopic cells, then the ectopic invagination is promoted by the increased lateral tension. The relaxation of the ectopic invagination happens if the external force disappears at the onset of the simulation (Figure \ref{fig:10}A), due to the competition between the mesoderm and ectoderm. The mesoderm invagination tends to be inhibited if the external force exists longer. (Figure \ref{fig:10}B, C) The abnormal apical constrictions still occur at both sides of the mesoderm boundary. As previously discussed, the combined twist production term mediated by a Hill function (Equation \ref{eq:12}) leads to a dome-shaped force distribution around the ventral midline. Cells at the boundary of that “dome” feel higher force gradients and deform the most. The mechanical feedback term promotes those cells to contract first. That’s why we observe mesoderm bending in the normal invagination, while in the case with local ectopic compression, the abnormal apical constriction happens instead without the inward curvature.
\section{Conclusion}
We proposed a vertex model to describe the mesoderm invagination mathematically. Our model considered the active forces contributing to the energy terms with line tension, and the passive mechanical response is achieved by the area elasticity of the yolk and cells. First, we examined our model with given sets of static tensions. It turns out that, in our model, the lateral tension is required to promote radial shortening and determine the depth of the furrow, which coincides with the conclusion of Conte’s model \cite{conte2012biomechanical} and the experiment \cite{gracia2019mechanical} Our model can successfully invaginate with active forces governed by prescribed dynamics. And we suggest that apical constriction, basal expansion, and apicobasal shortening are necessary for the formation of a tubular furrow. Different from Polyakov’s model \cite{polyakov2014passive} where the apicobasal shortening results from the restoring force due to the mechanical property of the lateral side, in our model, the active lateral force mediates the apicobasal shortening more directly. We came up with a concise regulatory network to describe the dynamics of active tensions. We assumed that \emph{twist} is expressed in a graded fashion and affected by the mechanical deformation. In addition, \emph{twist} affects the myosin \RNum{2} intensity in the cell: positively on the apical and lateral side, negatively on the basal side. Our regulatory equations can reproduce the dynamics of apical tension gradient, basal tension decrease, and lateral tension increase, which are corresponding to apical constriction, basal expansion, and apicobasal shortening respectively. Then we integrated the regulation and the vertex model. The results show that our integrated model is also able to form a closed ventral furrow although there are anomalous deformations in the intermediate states. Our model also replicates the ectopic \emph{twist}  expression induced by mechanical compression. The ectopic expression of \emph{twist} fails the mesoderm invagination, which is consistent with the experiment \cite{farge2003mechanical}. Our model also predicts the ectopic invagination in presence of the external force gradient, which has not been verified in vivo. There are defects in the integrated model. For example, the mechanical term leads to a “dome-shaped” apical tension profile which has a smaller gradient around the ventral midline and a larger gradient away. This apical tension distribution results in mesoderm bending that deviates from the experimental observation. Also, unexpected deformations occur in the model under external mechanical constraints even though the regulatory network has already described the tension dynamics as expected. A possible reason could be that the mechanical information may not be equal-weighted as the genetic term in Equation \ref{eq:13} because it is not the primary source of \emph{twist} induction \cite{martin2020physical}. However, in our model, decreasing the weight of mechanical term fails to mechanically induce sufficient ectopic twist expression to inhibit the mesoderm invagination (data not shown). Therefore, there could be a novel way to consider the mechanical response in the reaction-diffusion equations. As for the regulation network, our model only contains one activator. This simplification may lose some features that could be critical to maintaining the tissue shape. Thus, a more complete regulation network should be considered in the future. Our attempt at the integration of the genetic regulation and the vertex model provides a relatively complete understanding of the mesoderm invagination process. But the accurate genetic regulation and mechanical feedback are still needed to be explored.
\bibliographystyle{apalike}
\bibliography{ref}
\end{document}